\documentclass{svproc}
\usepackage{url}
\usepackage{graphicx}
\usepackage{verbatimbox}
\usepackage{multirow}
\usepackage{array}
\usepackage{booktabs}
\usepackage{textcomp}
\usepackage[bottom]{footmisc}

\begin{document}
\mainmatter
\title{Exploring the Behavior of
\\Coherent Accelerator Processor Interface (CAPI) on IBM Power8+ Architecture and FlashSystem 900}
\titlerunning{CAPI on FS900}
\author{Kaushik Velusamy \and Smriti Prathapan \and Milton Halem}
\authorrunning{Kaushik Velusamy et al.} 
\tocauthor{Kaushik Velusamy, Smriti Prathapan, Milton Halem}
\institute{University of Maryland Baltimore County, MD USA,\break
\email{\{kaushikvelusamy,smritip1,halem\}@umbc.edu},  \break
\texttt{https://carta.umbc.edu/}}

\maketitle             

\let\thefootnote\relax\footnote{Accepted for publication at 2019 International Workshop on OpenPOWER for HPC (IWOPH’19) International Supercomputing Conference HPC Frankfurt, Germany.}

\begin{abstract}
The Coherent Accelerator Processor Interface (CAPI) is a general term for the infrastructure that provides high throughput and low latency path to the flash storage connected to the IBM POWER 8+ System. CAPI accelerator card is attached coherently as a peer to the Power8+ processor. This removes the overhead and complexity of the IO subsystem and allows the accelerator to operate as part of an application.  In this paper, we present the results of experiments on IBM FlashSystem900 (FS900) with CAPI accelerator card using the ``CAPIFlash - IBM Data Engine for NoSQL Software'' Library.  This library provides the application, a direct access to the underlying flash storage through user space APIs, to manage and access the data in flash. This offloads kernel IO driver functionality to dedicated CAPI FPGA accelerator hardware. We conducted experiments to analyze the performance of FS900 with CAPI accelerator card, using the Key Value Layer APIs, employing NASA's MODIS Land Surface Reflectance dataset as a large dataset use case. We performed Read and Write operations on datasets of size ranging from 1MB to 3TB by varying the number of threads.  We then compared this performance with other heterogeneous storage and memory devices such as NVM, SSD and RAM, without using the CAPI Accelerator in synchronous and asynchronous file IO modes of operations. The asynchronous mode had the best performance on all the memory devices that we used for this study. In particular, the results indicate that FS900 \& CAPI , together with the metadata cache in RAM, delivers the highest IO/s and OP/s for read operations. This was higher than just using RAM, along with utilizing lesser CPU resources. Among FS900, SSD and NVM, FS900 had the highest write IO/s.  Another important observation is that, when the size of the input dataset exceeds the capacity of RAM, and when the data access is non-uniform and sparse, FS900 with CAPI would be a cost-effective alternative.
\end{abstract}

\section{Introduction}\label{sec:introduction}

Emerging needs as we approach for exascale computing bring new challenges related to heterogeneous computing which requires intensive analysis of massive and dynamic data . These new data challenges require us to move from conventional processing to new paradigms to better exploit and optimize the IO system architecture to support these needs \cite{BigData}. The increase in computational bandwidth brings a massive increase in memory bandwidth requirements. High bandwidth memory (HBM) \cite{HBM2} enabled FPGA accelerators address these challenges to provide a large memory bandwidth. The use of accelerators as co-processors is the most prevalent way to achieve performance gains to compensate for the slowdown of Moore’s Law. However, there are limiting factors in this approach such as 1) device driver overheads 2) operating system code path length 3) and CPU overhead required in managing the IO requests. Coherent Accelerator Processor Interface (CAPI) was developed to address these challenges, where an accelerator is attached coherently as a peer to the CPU through its IO physical interface\cite{stuecheli2015capi}. 

This method significantly increases the performance of the accelerator when compared to the traditional IO model. The performance of a CAPI attached heterogeneous storage device is drastically improved by reducing the overhead of the device driver in the operating system. It also allows direct memory access to the underlying storage without the calls to the device driver and without any intervention from the operating system thereby, reducing the code path length.

Our goal is to understand and analyze the performance capabilities of Power8+ CAPI infrastructure through FlashSystem900 (FS900) and compare it with the other heterogeneous memory devices without CAPI. For this experiment, we use the Key Value (KV) Layer (ArkDB) APIs provided by IBM CAPIFlash library \cite{software-lib}. The CAPIFlash library provides a multiple of 4K  block access to the flash. Our benchmark application uses this facility provided by the library to read a 4K block for the values. Our benchmark results directly relate to real-world application performance. Apart from obtaining the peak performance (IO/s, OP/s and Time) of the heterogeneous storage devices, we also focus on storage and computational efficiency.

\par This paper makes the following contributions:

\begin{itemize}
  \item Analyze the performance of the CPU threads for read / write workload on IBM FS900 with CAPI and compare it to RAM, NVM and SSD (Without CAPI) using the CAPIFlash library's synchronous and asynchronous modes of communication.
  \item Analyze the time taken to completion with varying threads, optimal IO/s (Input Output per second) and OP/s (OPerations per second) of heterogeneous storage devices.
  \item Analyze the impact of CPU usage in handling the IO requests in heterogeneous storage systems when using CAPIFlash library.
  \item Provide a performance evaluation of IBM FS900 with CAPI and RAM without CAPI when the input dataset size exceeds the capacity of RAM (1 TB).
\end{itemize}

The paper is organized as follows. Section 2 presents the background, featuring the enhancements that adding CAPI to the architecture brings. Section 3 details the library used in our experiments - CAPIFlash Library Design, from the kernel to the API. Section 4 describes the Test Bed and System Configuration. Section 5. explains the Experimental Setup and the Benchmark. The Performance Results are presented in Section 6. Related works are discussed in Section 7, followed by the conclusions and the summary in Section 8.

\section{Background}\label{sec:background}

The main problem we encounter with traditional flash storage is the CPU overhead leading to stranded IOs \cite{SolutionRefGuide}. There was a CPU overhead in managing the IO requests as it flows through the CPU to the storage device. Even with the multi-core architecture, more CPUs are needed to handle the IO request. This results in the wastage of significant portions of the available CPUs that are tied to the IO overhead, rather than doing the useful computations. For applications that require high computational power, multi-core processors often do not suffice to provide high performance. The use of hardware accelerators such as GPUs, ASICs, and FPGAs can provide increased performance for massively parallel programmable architectures for a wide range of applications. 

On each Power8+ processor, there is one CAPI interface called Coherent Accelerator Processor Proxy (CAPP).  The PCIe Host Bridge (PHB) on the processor connects to the PCIe IO links. The CAPP unit together with PHB acts as memory coherence, data transfer, interrupt, and address translation agents on the Symmetric Multi Processor (SMP) interconnect fabric for PCIe-attached accelerator  \cite{7029173}. This FPGA accelerator has a Power Service Layer (PSL) which provides address translation and system memory cache for the Accelerator Function Unit (AFU) and is connected to the Power8+ processor chip by the PCIe link. The combination of PSL, PCIe link, PHB, and CAPP enhance the capabilities of AFUs. AFUs operate coherently on the data in memory, as peers of other caches in the system.

An application which runs on the processor core has a virtual address space on the memory for performing IO with other devices attached to the processor. In a heterogeneous compute cluster without CAPI, a hardware accelerator such as an FPGA is attached using a PCIe. This leads to having many separate copies of the data, thereby adding overhead to using the FPGA. Since the copies of the data reside in different memory address spaces, any changes that were made to the data by the application would not be coherent with that on the FPGA.

With CAPI FPGA attached to a PCIe, the PSL layer is used to access shared memory regions and cache areas as though they were a processor in the system. This ability enhances the performance of the data access and simplifies the programming effort to use the storage device. Instead of treating the hardware accelerator as an IO device, it is treated as a co-processor, which eliminates the requirement of a device driver to perform communication and the need for Direct Memory Access that requires system calls to the operating system (OS) kernel. By removing these layers, the data transfer operation requires much fewer clock cycles in the processor, improving the IO performance~\cite{s822lc}.

Hence with CAPI, there are fewer overheads in IO, thereby minimizing thousands of instructions. All the data is coherently managed by the hardware. Furthermore, an FPGA accelerator can now act as an additional core in the server with a coherent memory. CAPI enables many heterogeneous memory devices to coherently attach to the Power8+ processors, facilitating an environment to connect various high bandwidth IO devices~\cite{CAPI-ibm}. By using CAPI, terabytes of the flash storage array can be attached to the Power8+ CPU via an FPGA. CAPI enables  applications to get direct access to the hardware storage (a large flash array) with reduced IO latency and overhead, thereby increasing the read / write performance compared to the standard IO-attached flash storage~\cite{SolutionRefGuide}. CAPI also eliminates the context switch penalties caused by interrupts. All these benefits enable CAPI to provide a hybrid computing environment.  For more information on the Coherent Accelerator Processor  Interface on  Power8+ Processor chip refer to the CAPI user manual \protect\cite{CAPI-Manual}.

\section{CAPIFlash Design}\label{sec:design}
In the Linux kernel, the Coherent Accelerator Interface (CXL) is designed to allow the coherent connection of accelerators (FPGAs and other devices) to a Power processor~\cite{Gilge2013RedpaperFO}. Through LibCXL, the user-space applications can directly communicate to a device (network or storage) bypassing the typical kernel/device driver stack. The CXL flash adapter driver enables direct access to flash storage for a userspace application. Applications which need access to the CXL Flash from the user space should use the CAPIFlash library. More information about the interfacing between CXL and the CAPIFlash is provided in ~\cite{CAPI-linux}.

The CAPIFlash library IBM Data Engine for NoSQL – Integrated Flash Edition~\cite{software-lib} was built on Power8 systems with CAPI. This library helps to create a new tier of memory by attaching up to 57 terabytes of auxiliary flash memory to the processor. This library provides two sets of public APIs for reading and writing to the physical address space on the flash device: 1. Cflash - Block Layer APIs and 2. ArkDB - Key Value (KV) Layer APIs. Our benchmark is focused mainly on the ArkDB KV layer APIs. The ArkDB KV layer API provides synchronous and asynchronous read/write requests to the flash memory. 

The send and receive operations to the KV Store (ArkDB) on the intended device (FS900, RAM, NVM, SSD) can be either synchronous or asynchronous. In synchronous operation, with a read/write request, the operation is initiated, following which the process is blocked and the system waits for the completion of the process. During this time, the ArkDB threads store and retrieve the data from the KV database instance (ArkDB) on the intended devices through CAPI.  

In an asynchronous operation, the processes run in non-blocking mode. It initiates the operation and does not wait for the completion to start the next operation. The caller would discover the completion of the operations later by polling the ArkDB.  Since the processes are non-blocking in asynchronous message passing setup, some computations can be performed when the message is in transit, thereby allowing more parallelism. 

\section{Test Bed and System Configuration}\label{sec:testbed}

In this work, the following storage devices have been used for performance analysis using the CAPI infrastructure. Flash memory, as a storage technology, is available in multiple forms, such as IBM FS900 or PCIe NVMe-SSD or a standard SSD product with a hard disk form factor. 

\subsection{CAPI Adapter}
\par The PCIe3 LP CORSA CAPI fibre channel Flash Accelerator x8 adapter (FC EJ16; 04CF) (Nallatech 385A72 with Intel Altera) FPGA accelerator card (with two 4GB DDR3), acts as a co-processor for the Power8+ processor. This is designed to offload CPU access to external fiber channel flash storage \cite{IBM-IO-Manual}. The adapter requires a direct-attach, point-to-point 8Gb fiber channel link to external storage, such as IBM FS900 \cite{IBM-IO-Manual}. This FPGA accelerator card includes the CAPI PSL, AFU  \protect\cite{CAPI-Manual} and interfaces to fiber channel IO ports to allow direct memory access to an IBM Flash System~\cite{SolutionRefGuide}. The EJ1K CAPI flash accelerator leverages the ability to provide high throughput, low latency connection to flash memory to address the scaling problems found in typical flash deployments~\cite{SolutionRefGuide}. The clock rate of the PCIe bridge is 33 MHz.

\subsection{IBM FlashSystem 900}
FS900 uses FPGA based flash arrays without involving processors and is cost effective when compared to DRAM~\cite{Gilge2013RedpaperFO}. The flash memory in the IBM FS900 has higher performance due to hardware only data path whereas a traditional SSD based flash memory is typically limited by the software processing. The IBM FS900 is connected to the CAPI accelerator through a fiber channel. This storage device is configured at RAID 5 and has a usable  capacity of 20 TB.  It also enables a distributed random-access memory and allows massive data parallelism. FS900 has persistent memory whereas RAM does not. The maximum 4KB IOPS 100\% random read is 1,100,000 and 100\% random write is 600,000.

We used one CAPI accelerator card, which has two ports, with the World Wide Port Name (WWPN) for each port mapped to a single volume. The KV experiments on the IBM FS900 were performed by setting the two volumes of FS900 (each 10 TB) in superpipe mode. This enables hardware acceleration,with the CPU offloading IO to the device.

\subsection{NVM and SSD}
\par NVMe is a host controller interface and a storage protocol, designed to accelerate the speed of data transfer using the PCIe bus, through processor-based storage solutions. It can read/write NAND flash memory to deliver the full potential of non-volatile memory in PCIe-based solid-state storage devices. The NVM in Power8+ system uses PCIe3, 3.2 TB NVMe Flash x4 adapter CCIN non-volatile memory controller (HGST Inc Ultrastar SN100) series with Non-Volatile Memory Express (NVMe) SSD. For the PCIe Interconnect, each POWER8 processor has 32 PCIe lanes running at 9.6 Gbps full-duplex. The theoretical bandwidth is: 32 lanes × 2 processors × 9.6 Gbps × 2 = 153.6 GB/s~\cite{s822lc}. 
\par The Samsung SATA SSD (MZ7LM3T8HCJM) used in the S822LC system has a capacity of 3.840 TB with a form factor of 2.5" inches. This SSD use a V-NAND technology and has a data transfer rate of 600 MBps. There are 2 SSDs in this Power8+ system and are connected to the integrated SATA controller in the motherboard. 

\subsection{RAM}
\par RAM in the IBM Minsky system has a total capacity of 1TB. The Power8+ S822LC computing server provides 8 DIMM memory slots each with 128 GB, allowing for a maximum system memory of 1024 GB DDR3 ECC at 1333 MHz . Each Power8+ processor has four memory channels running at 9.6 Gb/s capable of reading 2 Bytes and writing 1 byte at a time. The total theoretical memory bandwidth is:  4 channels * 9.6 Gb/s * 3 Bytes = 115.2 GB/s per processor module~\cite{s822lc}.

\subsection{Processor}
This CAPI enabled system S822LC server has two 64-bit IBM Power8+ processors with 10 cores each with 8 threads/core, running at a clock rate of 4 GHz. This system runs Ubuntu OS (version - 16.04.5 LTS ) on the PPC64LE architecture. It has a 64KB D cache, 32KB I cache, 512KB private L2, 8MB L3 per core (96M)on a 22nm chip. The processor to memory bandwidth is 170 GB/s per socket i.e. 340 GB/s per system and IO bandwidth of 64 GB/s simplex. Each Power8+ processor has 2 memory controller. Each memory controller has a two buffer L4 cache and each of them are connected to four RAM DIMM slots~\cite{s822lc}. POWER8+ is a revised version of the original 12-core POWER8 from IBM. The main new feature is the support for Nvidia's bus technology NVLink, connecting up to four NVLink devices directly to the chip. IBM removed the ``A Bus and PCI interfaces" for SMP connections to other POWER8 sockets and replaced them with NVLink interfaces.

\section{Experimental Setup and Benchmark Overview} \label{sec:benchmarks}

\subsection{Dataset}
The input data used for this work is NASA\textquotesingle s MODIS ( MODerate-resolution Imaging Spectrometer ) Terra/Aqua Surface Reflectance data \cite{MOD}. Surface reflectance is the amount of light reflected by the surface of the earth. In addition to the geo-location coordinates, the data contains several fields collected every 5 minutes at 250m, 500m and 1km resolution as 8-bit, 16-bit or 32-bit signed float/integer types. A sample structure for a refined subset of this dataset is shown in Table \ref{tab:Data-fields}. The data attributes that are of interest for this study are: Key, Value4, Value5 and Value6 with each entity of size 8 Bytes with a total of 32 Bytes per row in the input data.  In our experiments, the dataset ranges from 10 million KV pair records (543 MB) to 100 billion KV pair records (3.1TB). 

\begin{table}
\caption{Data fields in the MODIS Surface Reflectance dataset}\label{tab:Data-fields}
\begin{center}
\begin{tabular}{r@{\quad}rlrlrlrl}
\hline
Key & Value 1 & Value 2 & Value 3 & Value 4 & Value 5 & Value 6  \\
\hline\rule{0pt}{12pt}
No:  & Latitude & Longitude & Day &  Band 1 & Band 2 & Band 3\\
\hline\rule{0pt}{12pt}

    1 & 80.647079 & 34.720413  & 67 & -28672 & -28672 & -28672	\\
    2 & 80.658272 & 34.457428 & 67 & -28672 & -28672 & -28672	\\
    .  & .. 		        & .. 		         & ..	& ..     		& ..     		& .. 			\\ 
\hline
\end{tabular}
\end{center}
\end{table}

\subsection{Performance Metrics}

Some of the performance metrics used for this experiment are as follows. \verb|OP/s|  is the number of Set(write)/Get(read) operations per second submitted by the CAPIFlash library. \verb|IO/s| or IOPS is the number of IO operations per second. \verb|Read time| / \verb|Write time| is the time taken to complete entire reading / writing of data (in seconds).

\subsection{CAPI Flash parameters}

The following are the important parameters used for evaluating this benchmark; 1. \verb|Hcount|, 2. \verb|Threads|, 3. \verb|Queue Depth|, 4. \verb|Metadata Cache|.

The CAPIFlash library provides access to the flash in multiples of 4K blocks. The \verb| hcount| (-h) parameter is a calculation of how the KV pairs fit into a physical 4K block \cite{software-lib}. This, in turn, provides a more balanced allocation of storage space, thereby increasing the performance and storage efficiency.  \verb|Hcount| is calculated by dividing 4K by ( 32 (Bytes per KV entry in the dataset) + 8 Bytes (metadata)). Hence, a value of 100 would give the best performance for a dataset where each record is of 32 Bytes. The value 100 in this calculation means that the read/write operations would be performed on 100 KV pairs in parallel in a given 4K block. The value of \verb|hcount| is calculated as the total number of KV pairs in the input data divided by 100 in this example. The total memory allocated in Bytes (\textit{inuse}) would then be close to the Bytes written to the KV instance (\textit{actual}), thus providing an optimal storage efficiency. 

There are two different threads with regards to this benchmark: 1. application threads and 2. ArkDB threads. Each read / write is an operation submitted by application thread(s) to the ArkDB, and then the application threads poll the ArkDB for their completion. There is a maximum number of operations that can be queued to the ArkDB. The maximum value for \verb|QueueDepth| (QD) is 1024 with a default of 400. The queued operations are then submitted by the ArkDB threads to the storage media. QD represents the number of ArkDB operations which may run in parallel. 

In synchronous mode, an application thread sends a command (read/write) and waits for completion while the ArkDB threads do the work. As a result, one thread can do QD=1. More threads are required to handle a greater value of QD. In asynchronous mode, each operation submitted adds to the QD. A certain QD is required to reach the maximum bandwidth of the card. In asynchronous mode, with more threads, more CPU is used. The threads can be  increased until the maximum performance is obtained. In asynchronous mode, one thread may submit any number of operations (QD=N) and then check for the completion of each command. The ArkDB threads do the database side of the work. Too many ArkDB threads increase CPU usage without increasing  \verb|OP/s| and reduce the overall performance. Hence finding an optimal thread for the peak performance is important. The threads mentioned in the remainder of the experiment are the ArkDB threads. The number of threads can be passed in as an input parameter which specifies the total ArkDB threads. 

FS900 can be used either with \verb|MetaDataCache|  turned on or off. This allows a part of the RAM to be used as a cache for the data access to the FS900. Note that, the number of IO/s for FS900 which are greater than 400K are from the ArkDB cache hits, which is RAM. In order to get the maximum performance for the benchmark, all FS900 experiments were run with ArkDB metadata cache hits (RAM) enabled unless specified. 

\subsection{Benchmark Application}

The benchmark application used, runs a simple read / write operation to the KV store (ArkDB) on the intended devices (FS900, RAM, NVM, SSD) using the KV layer APIs of the CAPIFlash library. Only FS900 uses the CAPI accelerator card in the real IO mode, whereas NVM and SSD do not use the CAPI accelerator and are configured in the file IO mode. When the benchmark application requests to run to a file on NVM or SSD, the Block API diverts to the file before reaching the CAPI card \cite{SolutionRefGuide} . When the benchmark application requests to run on RAM, then the ArkDB API diverts the calls to RAM, before reaching the CAPI card. The CAPI card is not used when running the application to RAM. FS900 experiments were run with enabling the  metadata cache (RAM). When an ArkDB operation gets a hit on the ArkDB metadata cache, the call does not go to the CAPI card or FS900. When the maximum bandwidth of a single Corsa CAPI card reaches 380K IO/s, any OP/s higher than 380K, are due to the ArkDB cache hits. We ran the KV R/W experiments by varying the threads from 1 to 128 and QD = 400 in both synchronous and asynchronous modes to find out the peak performance of each device.  Finally, we summarize the overall read/write performance for all storage devices. The CAPIFlash library version used in this experiment is 5.0.2706.

\section{Performance Results}\label{sec:performance}

\subsection{Performance of Write Operations}\label{sec:write}

\begin{figure*}
\centering
\begin{multicols}{2}
    \includegraphics[width=60mm,height=45mm]{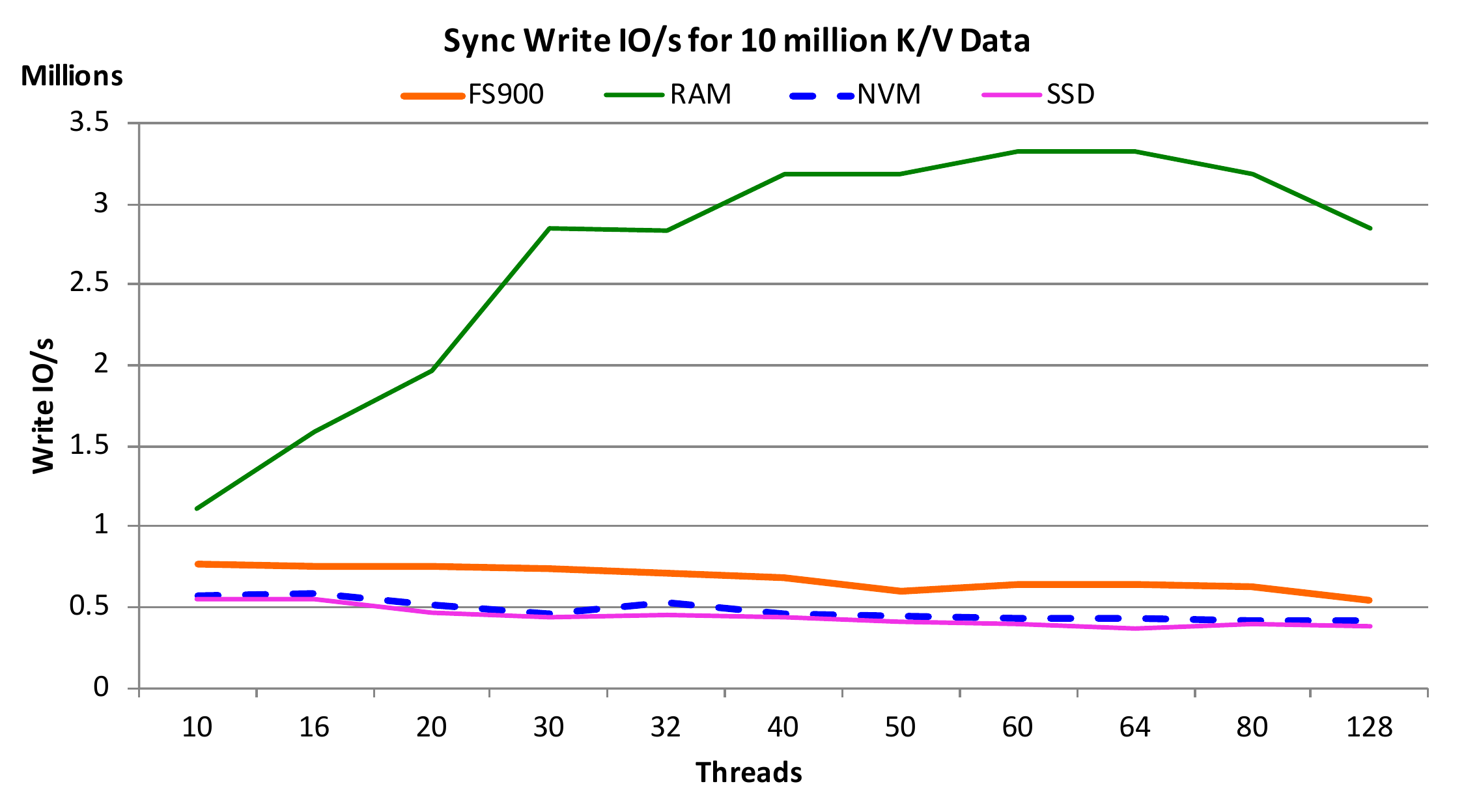}\par
    \includegraphics[width=60mm,height=45mm]{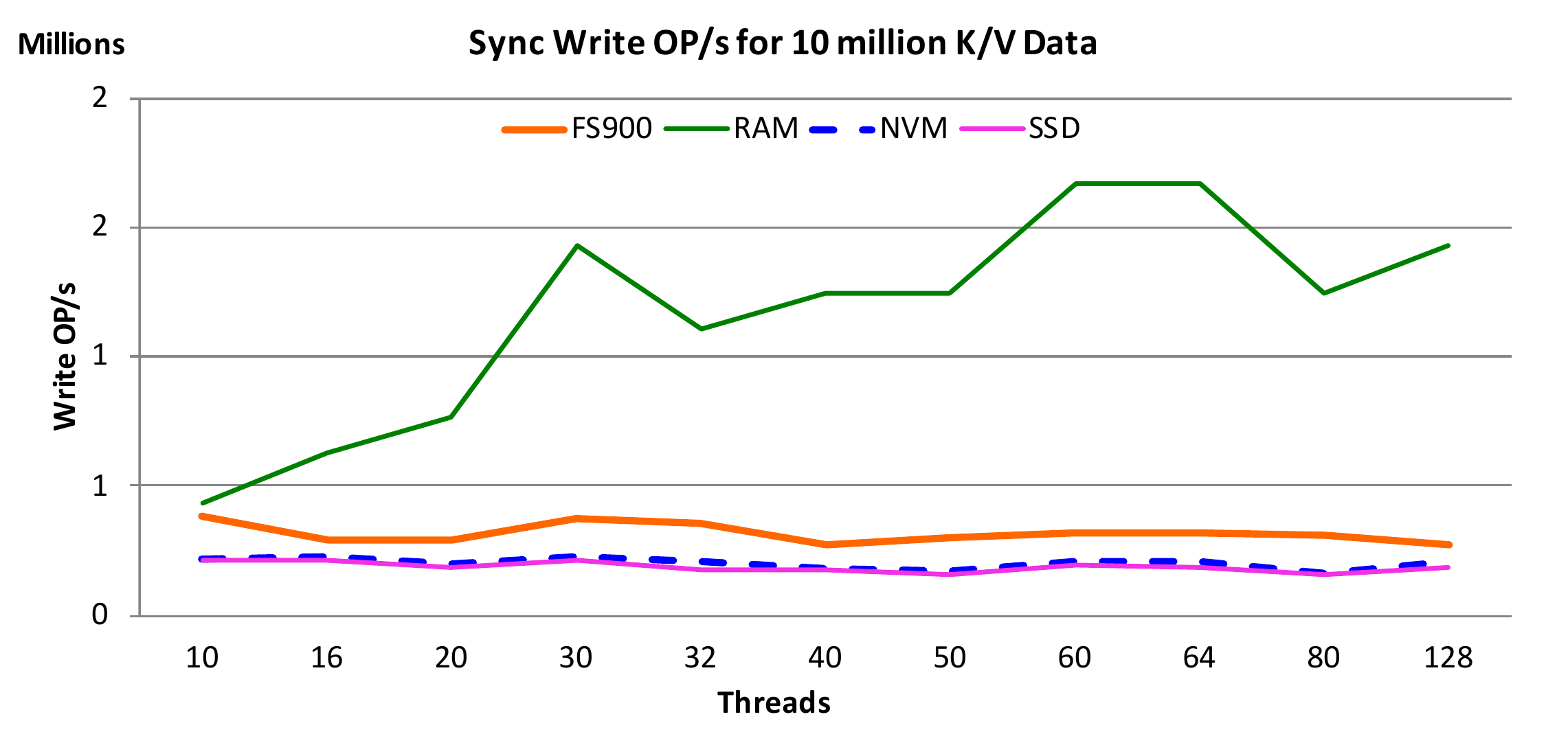}\par
\end{multicols}
\begin{multicols}{2}
  \includegraphics[width=60mm,height=45mm]{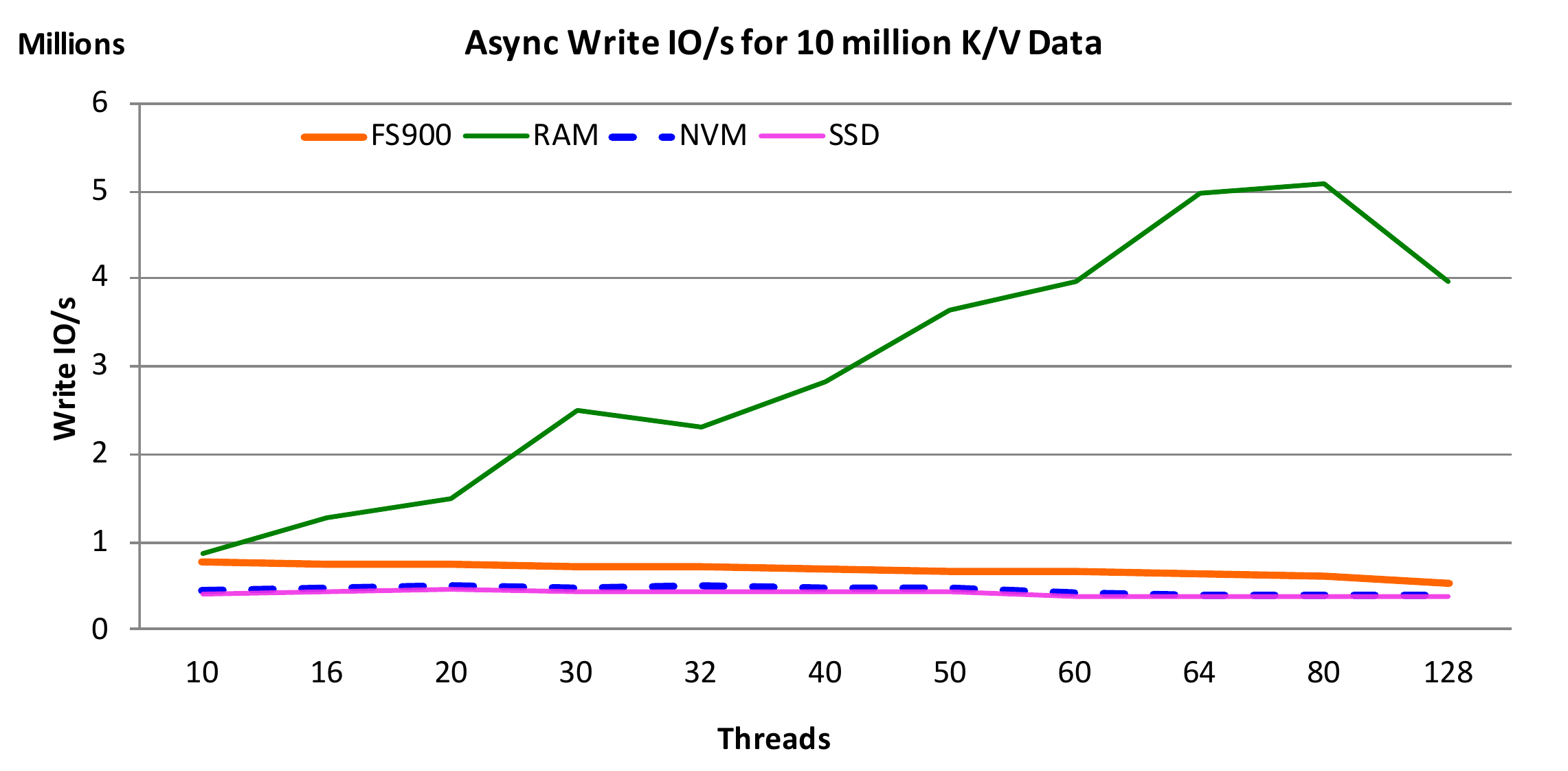}\par
  \includegraphics[width=60mm,height=45mm]{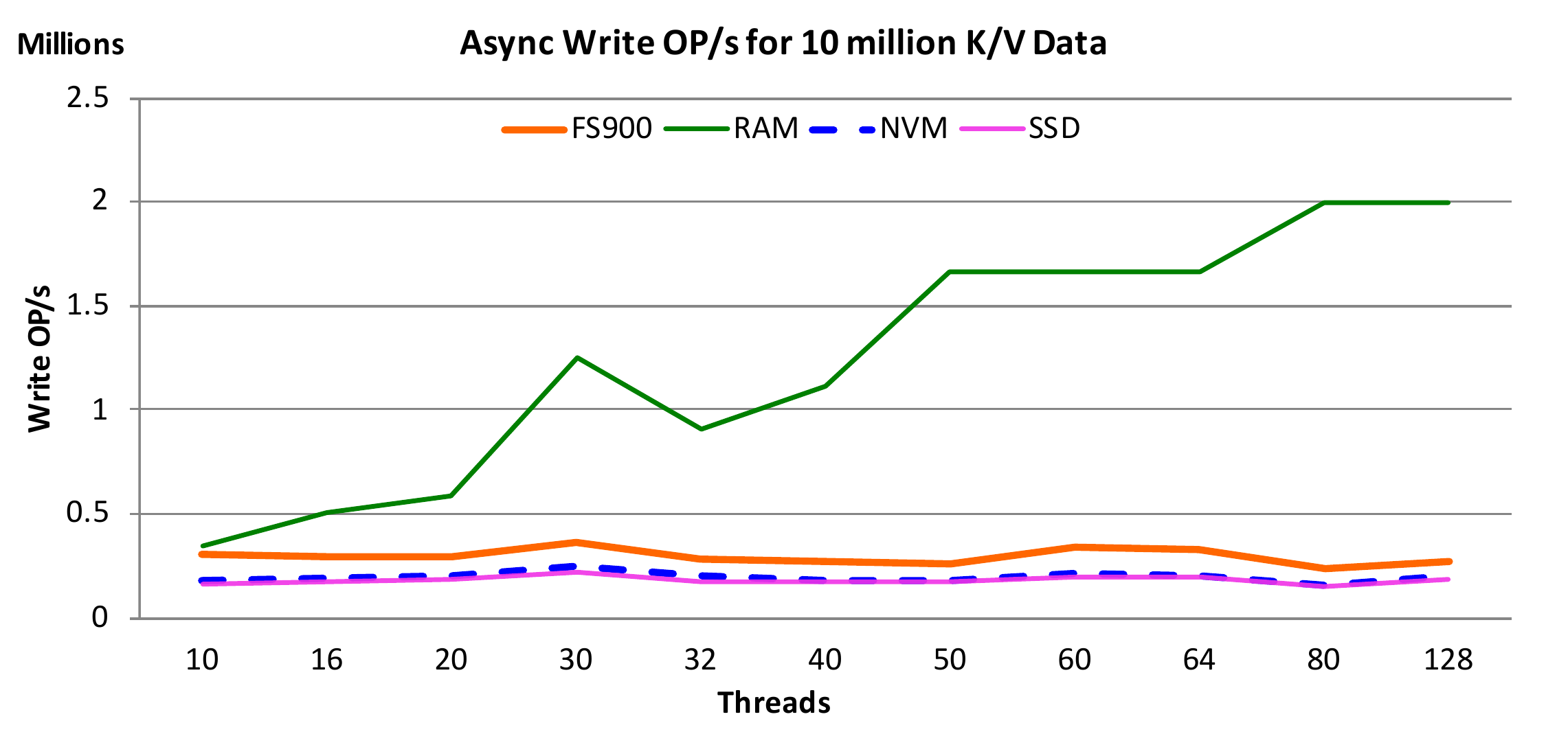}\par
\end{multicols}

\begin{multicols}{2}
\includegraphics[width=60mm,height=45mm]{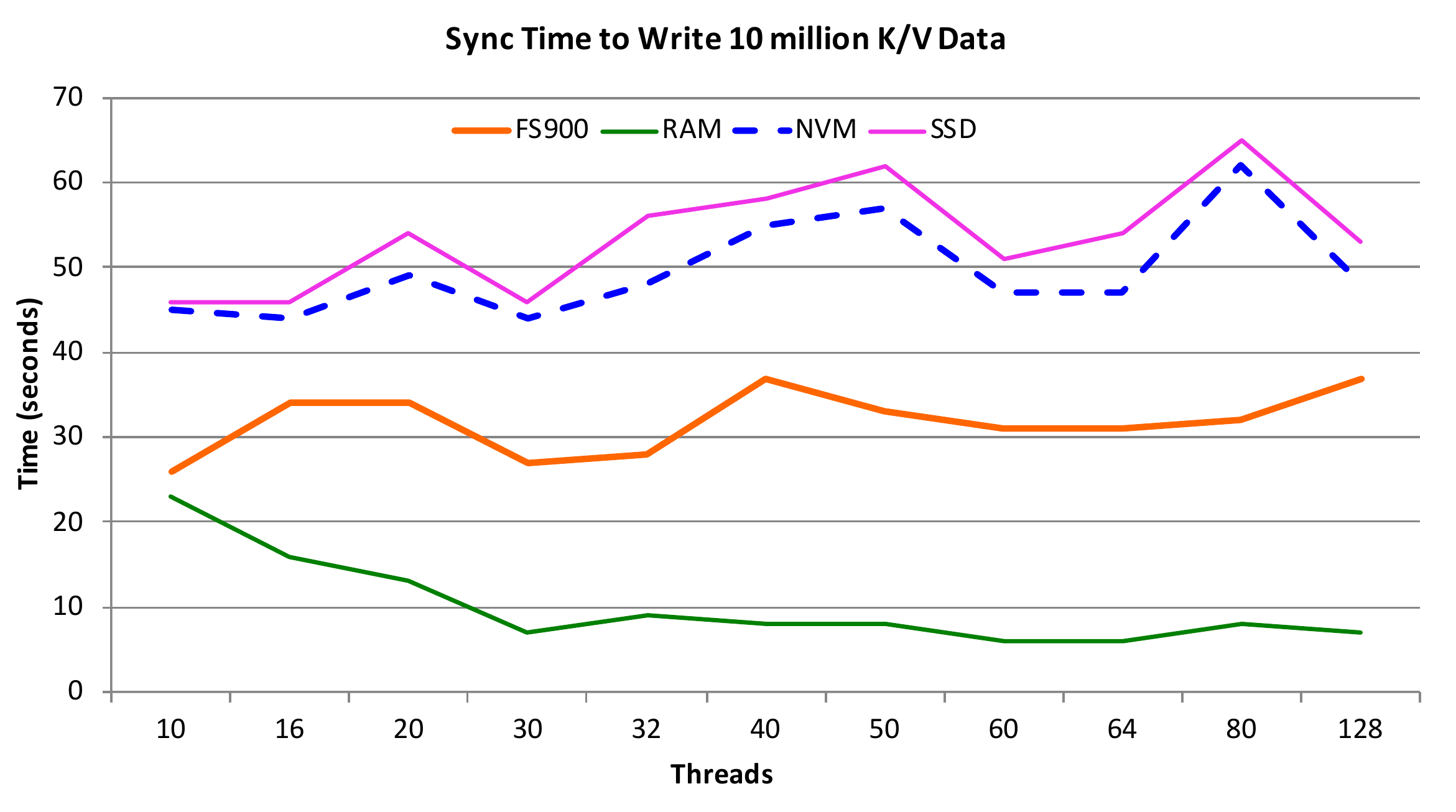}\par 
\includegraphics[width=60mm,height=45mm]{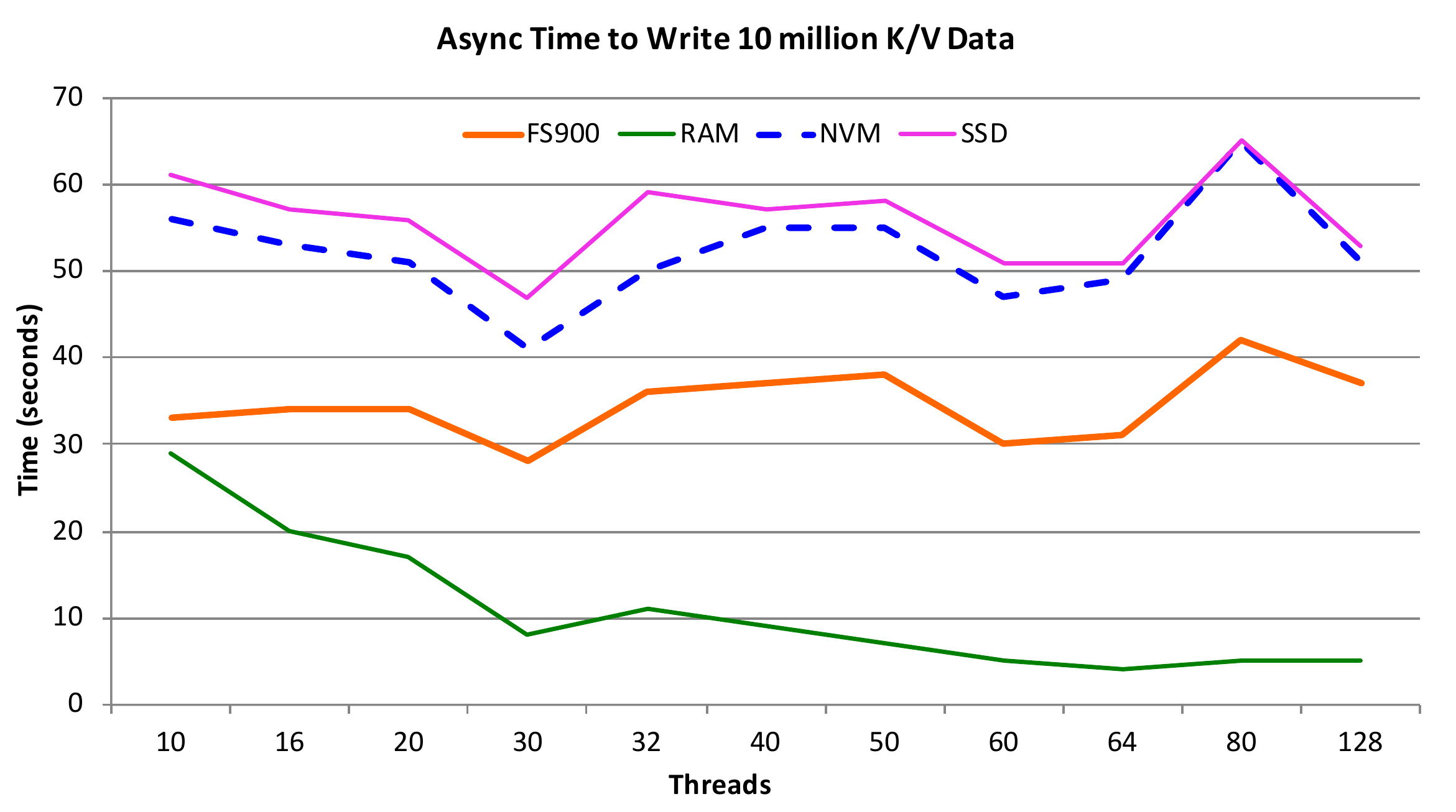}\par
\end{multicols}

\caption{Write performance for FS900, RAM, NVM and SSD}
\label{WCompare}
\end{figure*}

The results of Write performance of FS900, RAM, NVM and SSD for a dataset of 10 million KV pairs (0.5 GB) are in Figure ~\ref{WCompare}. In synchronous Write IO/s graph, for FS900, NVM and SSD, there was a plateau in the graph between 0.4 million to 0.8 million IO/s after 10 threads, with FS900 giving the maximum IO/s out of the three devices. For RAM, after 40 threads, the IO/s plateaued around 3.25 million after 40 threads. A similar trend was observed in the synchronous OP/s graph. With asynchronous Write IO/s graph, increasing the number of threads increased the IO/s performance in RAM until 5 million IO/s. There was little variation in the case of FS900, SSD and NVM. A similar trend was observed in the asynchronous OP/s graph. In the case of analyzing the time to completion for read/write operations, having a lower value is better. For write operations, the optimal threads for all devices were found around 30 threads. After 30 threads, each of the devices plateaued around a band of certain time regions without decreasing much time. From figure~\ref{WCompare} for write operations, it is clear that the best performance was found in the following order: RAM, FS900, NVM and SSD. Asynchronous mode was able to give the peak performance.  RAM consumed more numbers of threads to reach peak IO/s and OP/s, whereas for FS900, 10 threads were enough to reach the peak performance and lowest completion time.  The performance in asynchronous mode was found to be better since the computations do not block the process to wait for its completion, as opposed to  synchronous mode. In asynchronous mode, the ArkDB threads continue the computations while the messages are in transit and this approach allows more parallelism. The main advantage of FS900 with CAPI is the use of lesser CPU resources (if CPU threads are crucial to the application). Throughout the write operation, it is observed that using too many ArkDB threads decreases performance without increasing IO/s or OP/s and reduces the latency. For write operations, SSD and NVM showed similar performances as FS900. 
    
\begin{figure*}
\centering
\begin{multicols}{2}
\includegraphics[width=60mm,height=45mm]{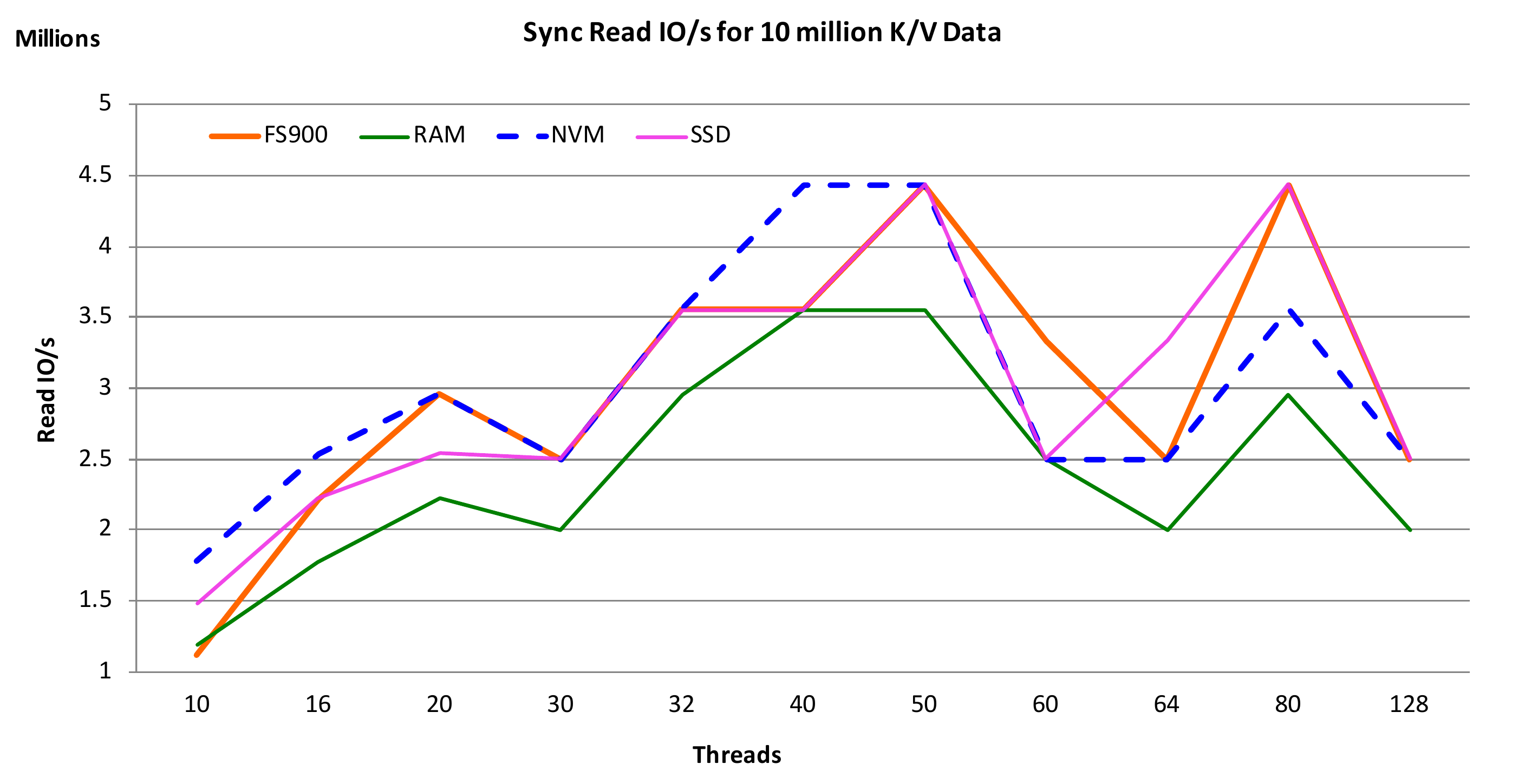}\par
\includegraphics[width=60mm,height=45mm]{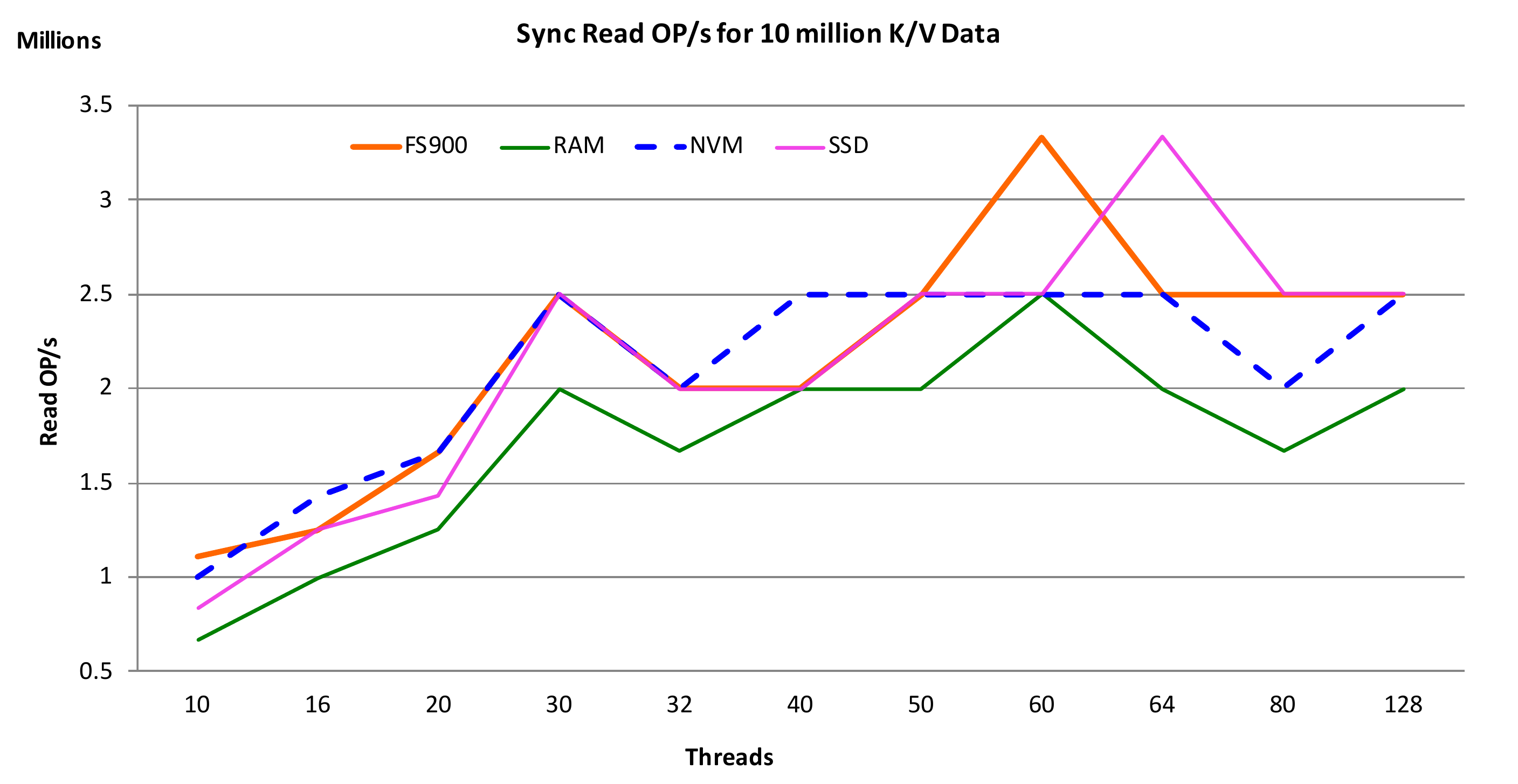}\par
\end{multicols}
\begin{multicols}{2}
\includegraphics[width=60mm,height=45mm]{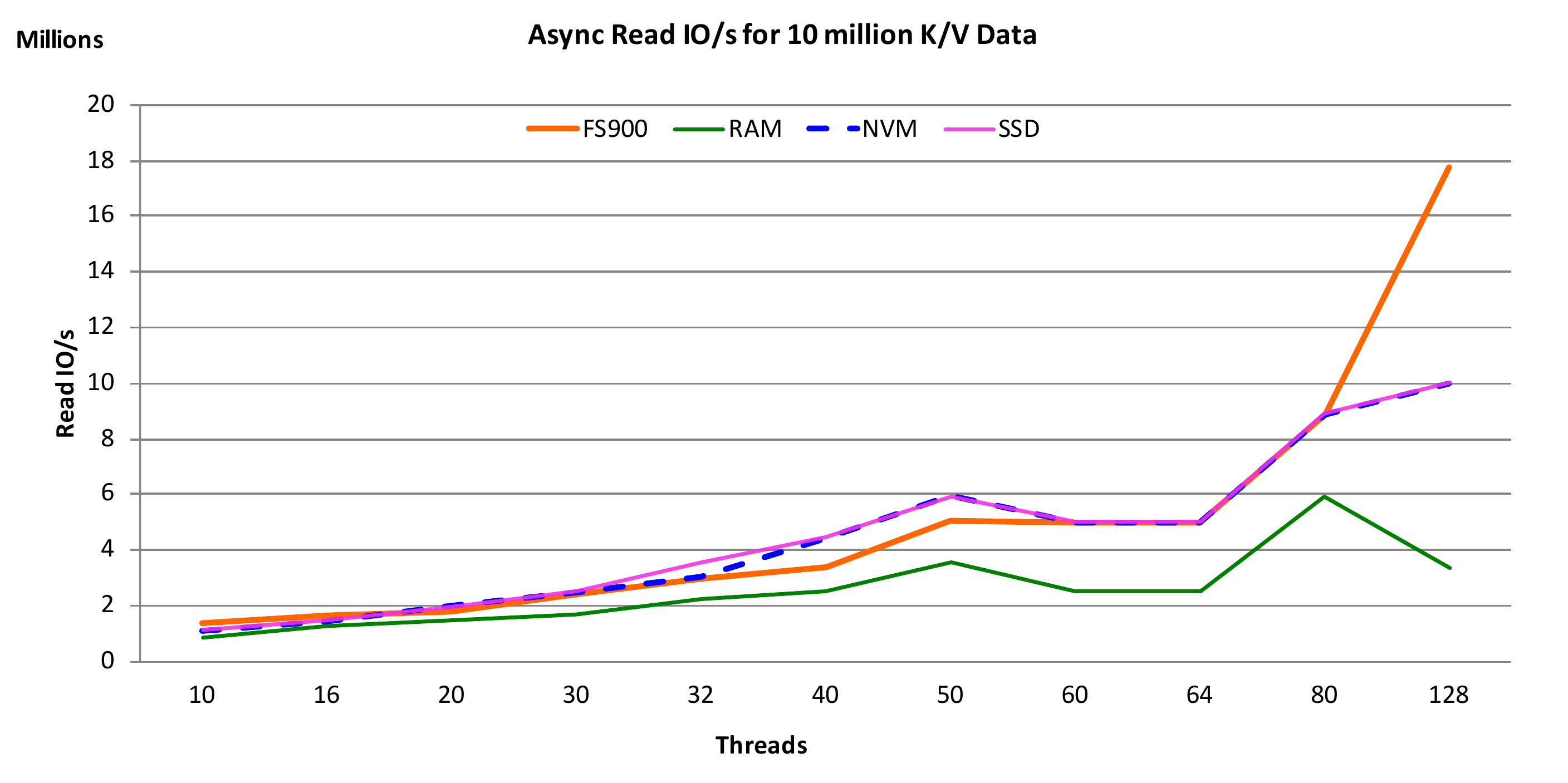}\par
\includegraphics[width=60mm,height=45mm]{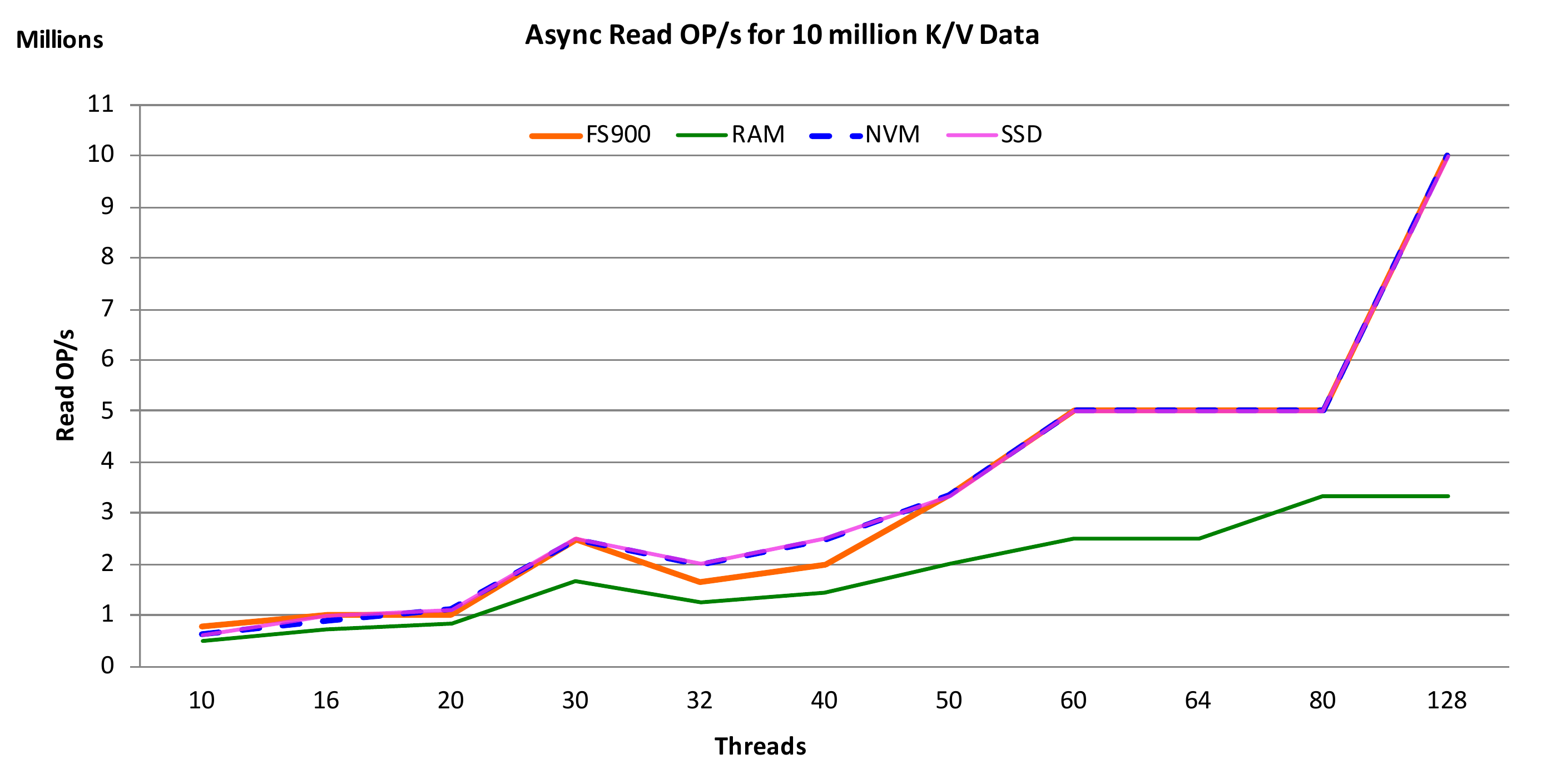}\par
\end{multicols}

\begin{multicols}{2}
\includegraphics[width=60mm,height=45mm]{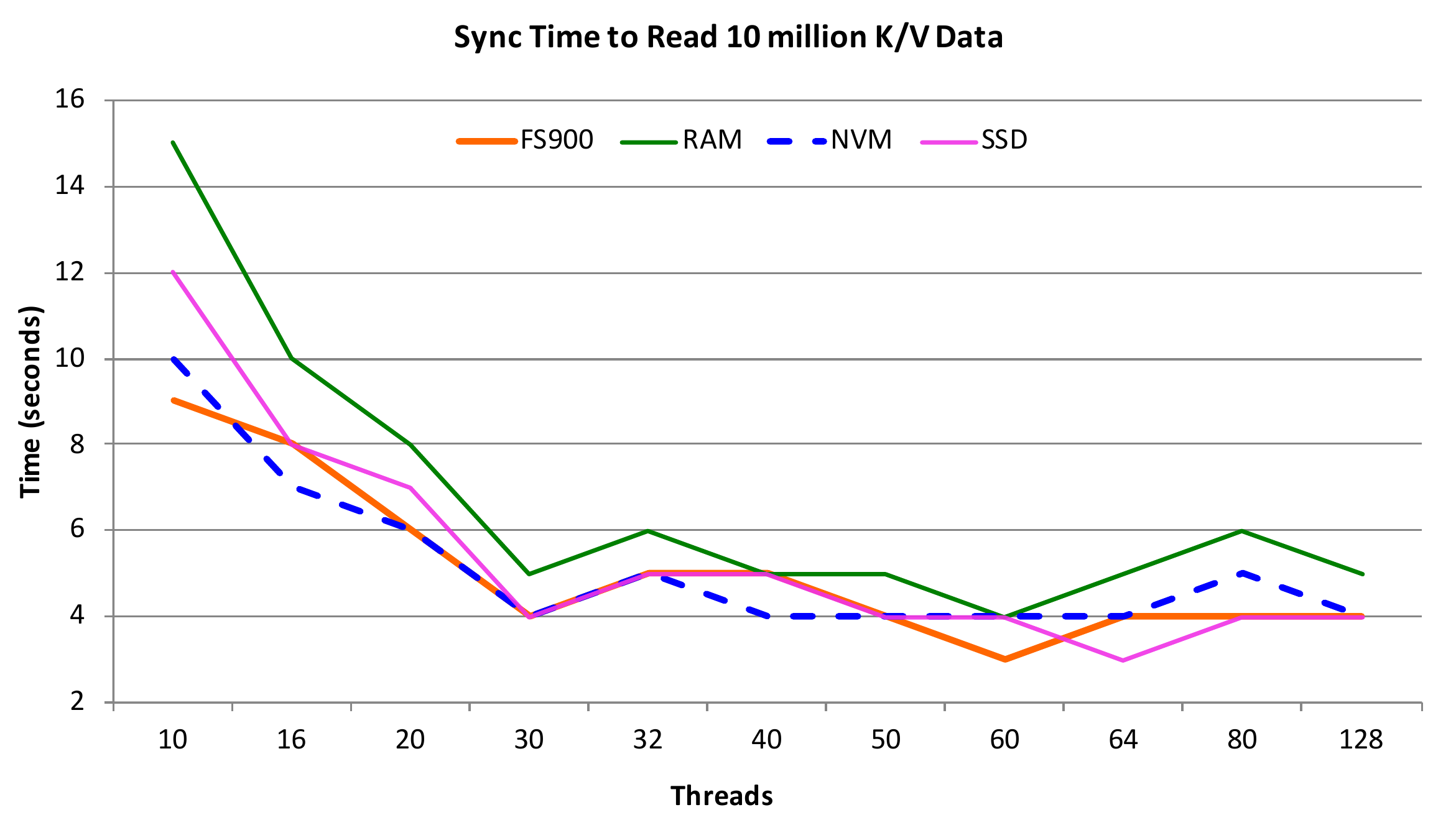}\par
\includegraphics[width=60mm,height=45mm]{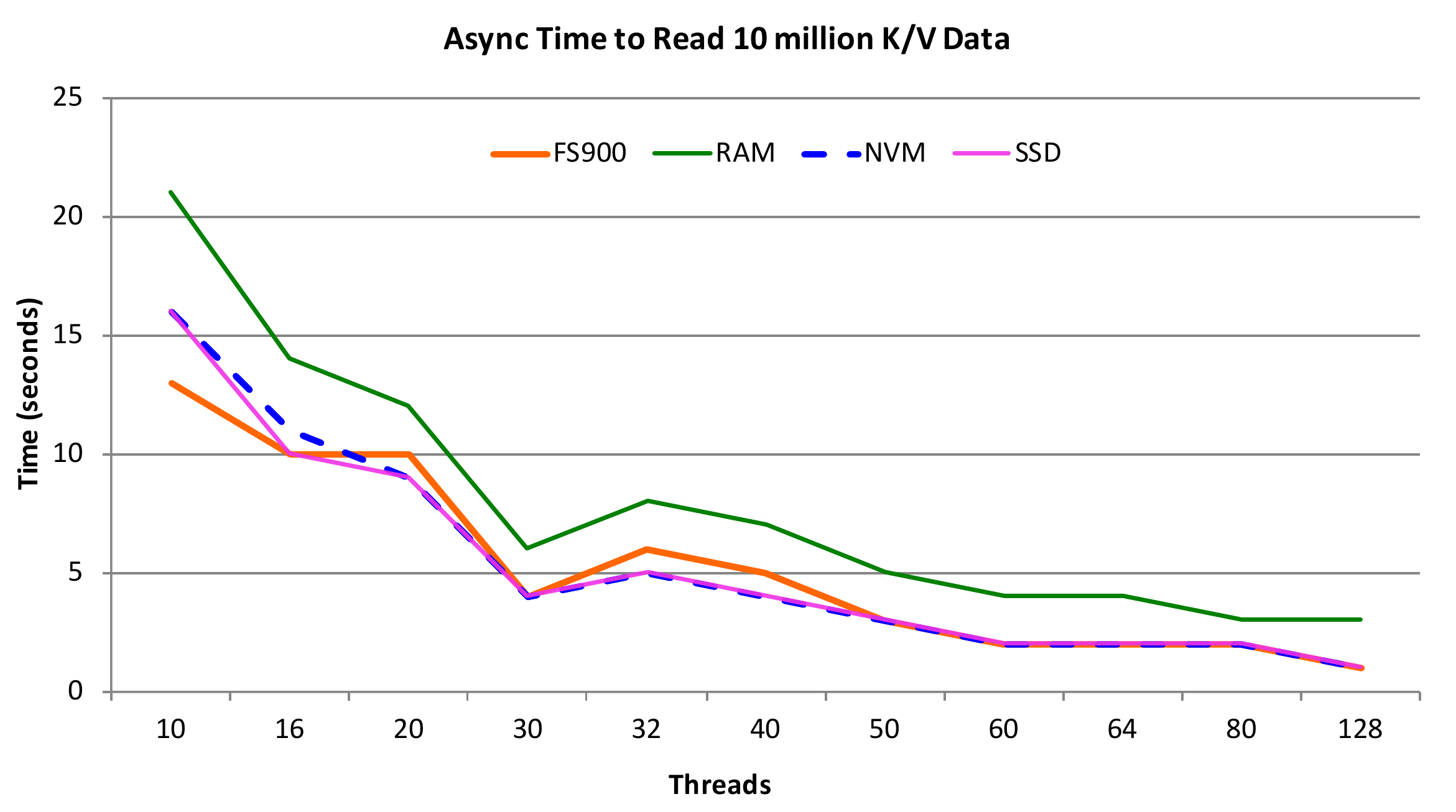}\par
\end{multicols}
\caption{Read performance for FS900, RAM, NVM and SSD}
\label{RCompare}
\end{figure*}

\subsection{Performance of Read Operations}\label{sec:read}

Figure ~\ref{RCompare} depicts the performance comparison on all the devices for reading 10 million KV pairs. For synchronous mode, the read IO/s for all devices lie around 1.5 million to 4.5 million, whereas in asynchronous mode, increase in threads almost gave a linear increase in read IO/s. In synchronous mode, FS900, NVM and SSD gave the maximum read IO/s compared to RAM. In the case of asynchronous, FS900 gave the maximum performance of 17 million IO/s. SSD and NVM at 10 million IO/s and RAM at 6 million IO/s.  For OP/s, FS900 performed better than RAM in both synchronous and asynchronous modes. FS900 took less time to complete the read operation in comparison with RAM. This was possible even with lesser number of threads. Based on our analysis, RAM is well suited for write-intensive workloads. For read operations, FS900 was found to be better than RAM. Having a RAM of large capacity comes at a higher overall cost. Similar or near to RAM performance could be achieved with FS900 at a lesser cost \cite{Gilge2013RedpaperFO}. CAPI facilitates attaching terabytes of the flash storage array to the Power8+ CPU with reduced IO latency and overhead compared to a standard IO-attached flash storage. FS900 with CAPI, when using the RAM metadata cache, performed 2x as many read operations in synchronous mode and 3x in asynchronous mode. The lower performance of RAM may be due to the CPU overhead required in managing IO operations. Without metadata cache, FS900 gave a steady 380K IO/s in both the modes consistently for all threads above 10.

\subsection{Summary of Read Write performance of RAM, FS900, NVM and SSD}\label{sec:summary-readwrite}
\begin{figure*}
\centering
	\includegraphics [width=80mm,height=35mm]{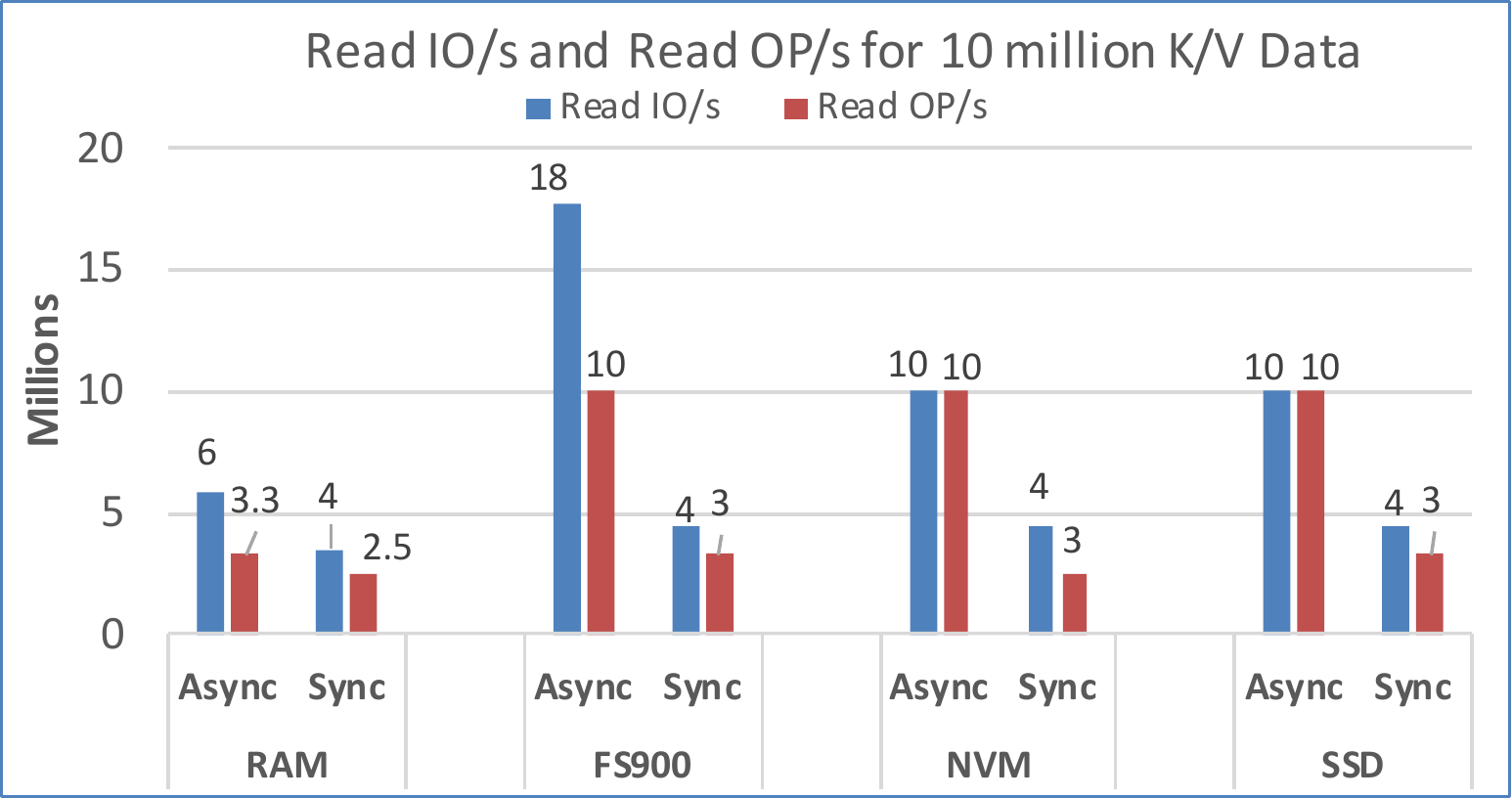}\par
	\includegraphics [width=80mm,height=35mm]{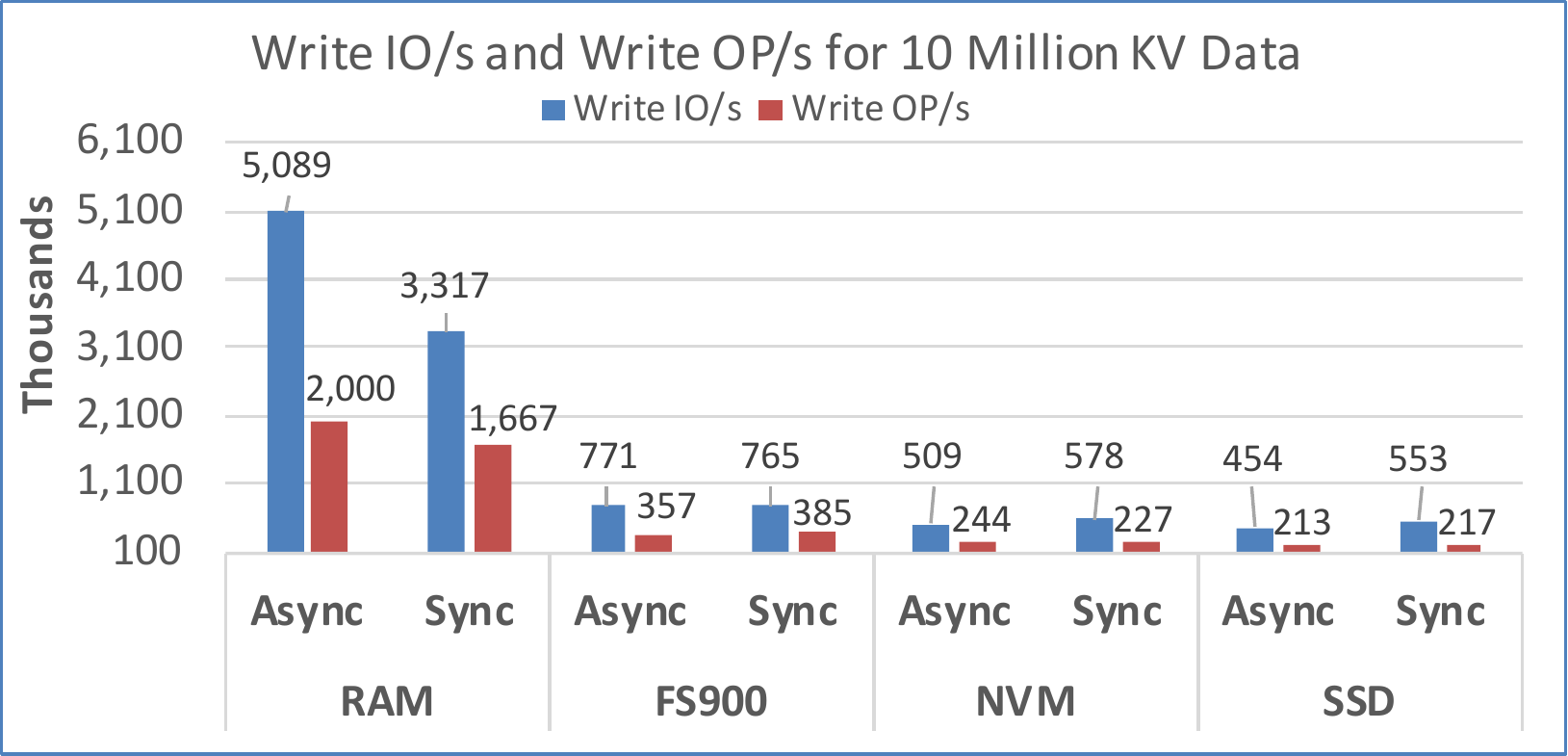}\par
    \includegraphics [width=80mm,height=35mm]{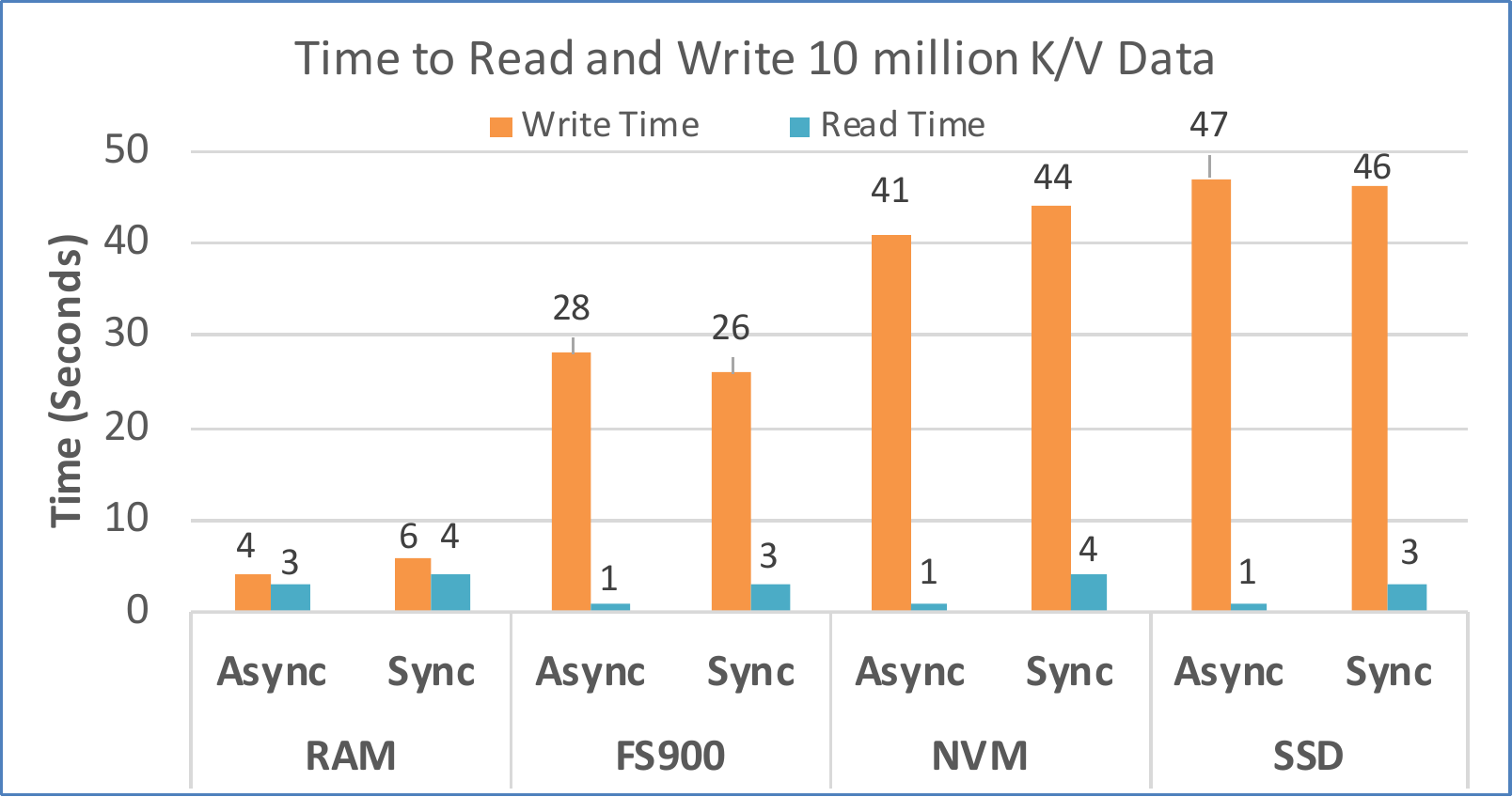}\par
\caption{Peak performance of FS900, RAM, NVM and SSD for 10 Million KV dataset}
\label{RWSummary}
\end{figure*}

The peak R/W performance for 10 million KV pairs on all the devices is summarized in Figure ~\ref{RWSummary}.
The best values of read and write OP/s, IO/s and time are compared for all the devices. RAM outperformed all the other devices for write in terms of OP/s, IO/s and completion time.  Hence for write-intensive applications, RAM would provide the best performance. Although RAM performed the best for write operations, FS900 provided the highest IO/s for reads.  To read 10 million KV pairs, FS900 had the peak IO/s and lowest read completion time. RAM was 2x slower than the other devices in asynchronous mode and 1.3x slower in synchronous mode. After 128 threads, the performance plateaued. 

\begin{figure*}
\centering
    \scriptsize
    \includegraphics[width=1\textwidth]{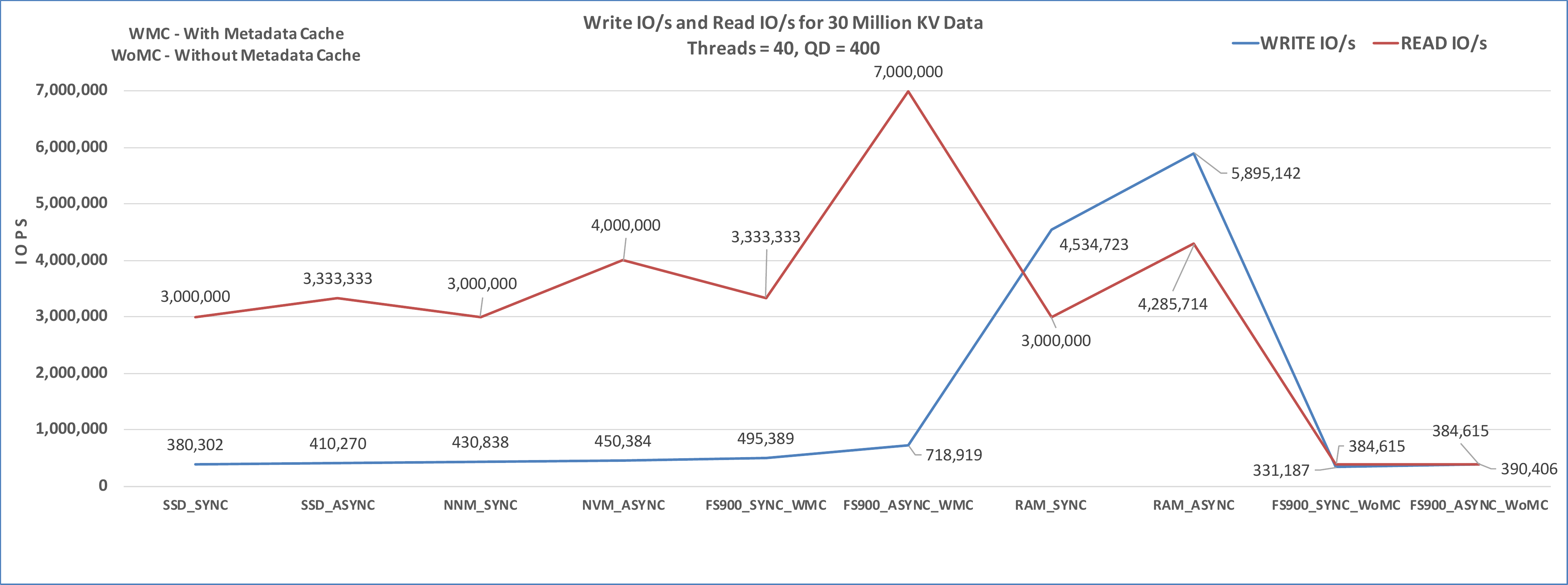}\par
\caption{IO/s performance of FS900, RAM, NVM and SSD for 30 Million KV dataset}
\label{setgetio}
\end{figure*}

\begin{figure*}
\centering
    \scriptsize
    \includegraphics[width=1\textwidth]{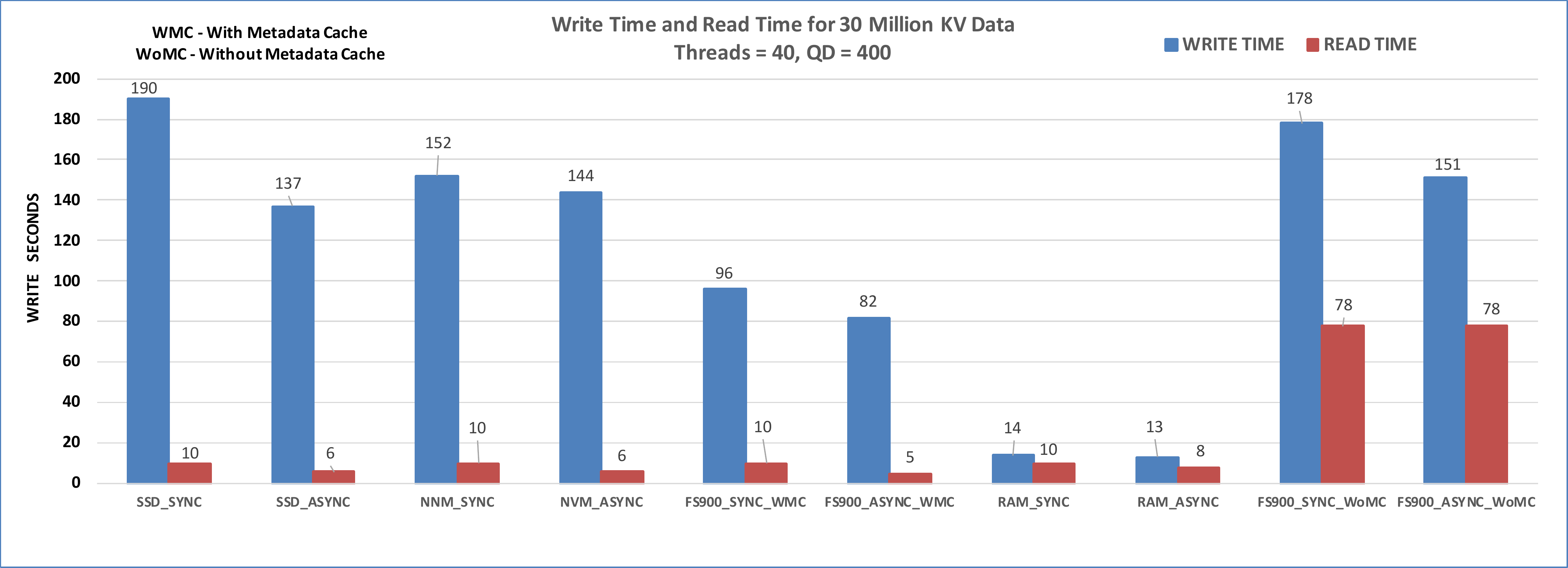}\par
\caption{Read Time and Write Time of FS900, RAM, NVM and SSD for 30 Million KV dataset}
\label{setgettime}
\end{figure*}

Based on the above analysis, we chose an optimal thread (40), QD (400) and compared the performance of synchronous and asynchronous modes on all devices for 30 million KV pairs (1.6 GB dataset) and the results are shown in Figure ~\ref{setgetio} and  ~\ref{setgettime}. Here, we also show the difference in the performance between FS900 with meta data cache  (WMC) enabled and without meta data cache (WoMC). For all the devices, asynchronous mode gave the peak performance and the lowest completion time, with FS900 giving the peak read IO/s compared to other devices. 

\begin{figure*}
\centering
\includegraphics[width=0.75\textwidth]{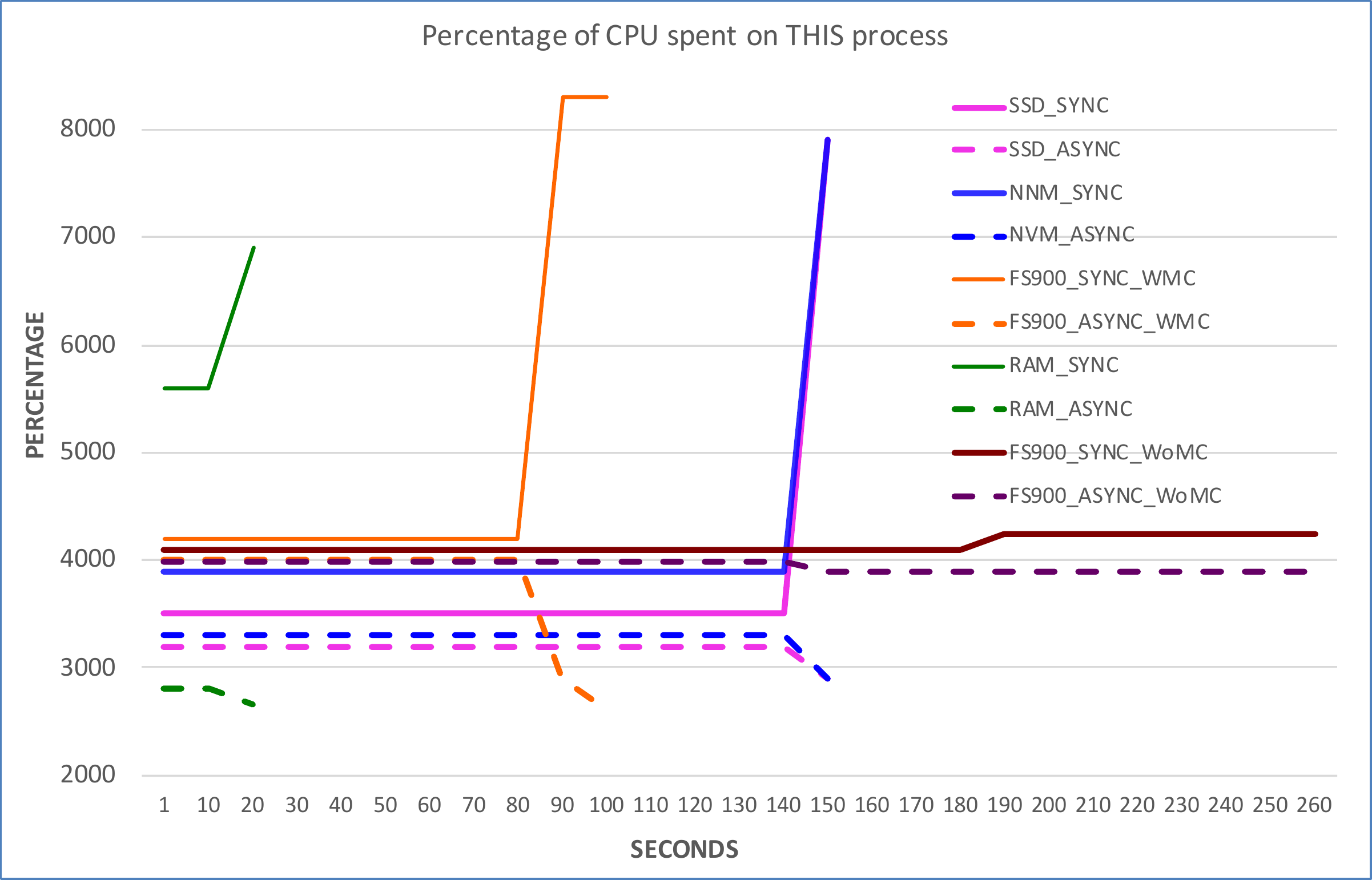}\par
\caption{  CPU\% used by the application process for 30 million KV dataset}
\label{CPU-usage}
\end{figure*}

\begin{figure*}
\centering
\includegraphics[width=0.75\textwidth]{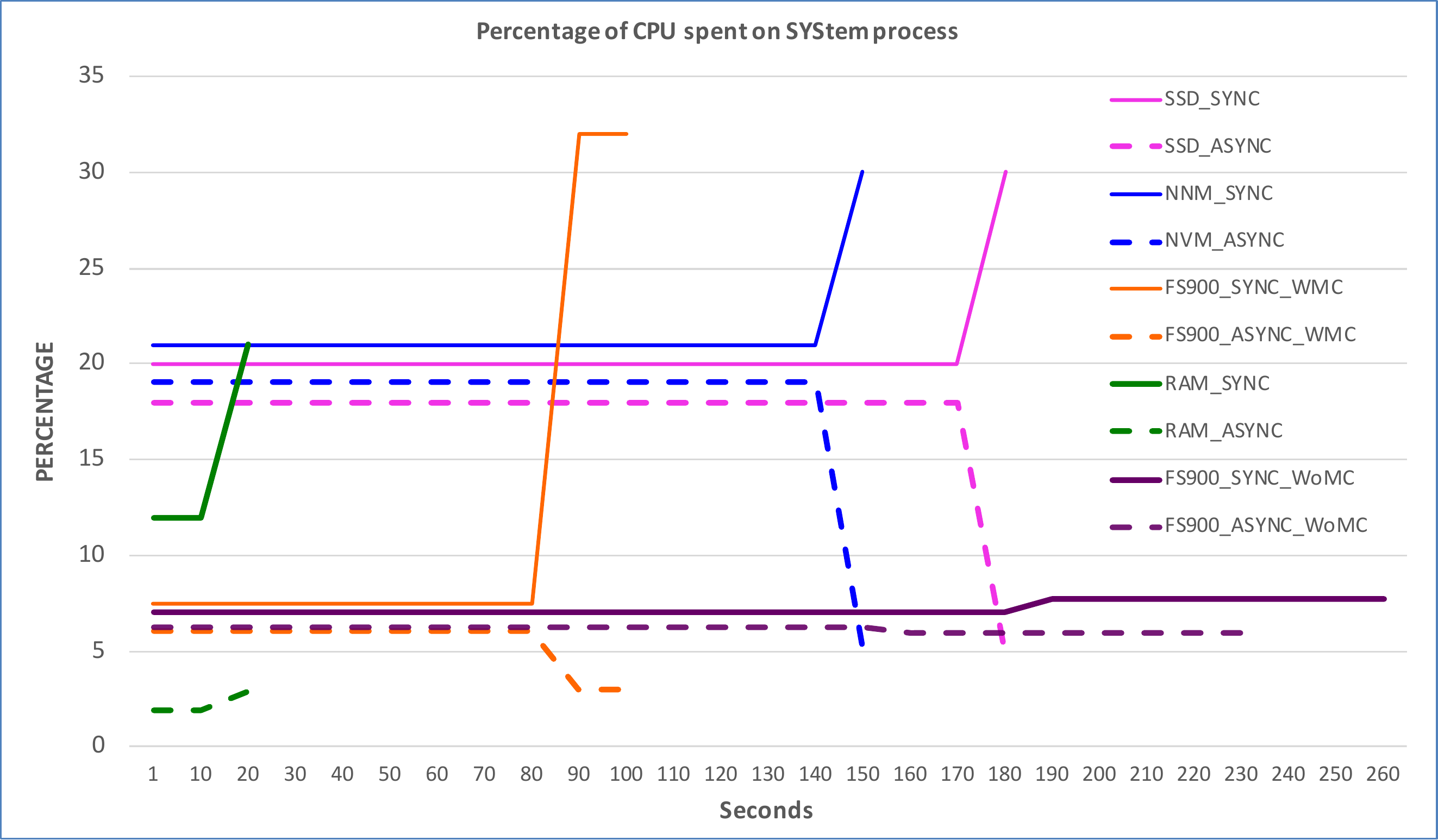}\par
\caption{ CPU\% used by the system process for 30 million KV dataset}
\label{SYS-usage}
\end{figure*}

Figure ~\ref{CPU-usage}  \& ~\ref{SYS-usage} details the CPU utilization (\% CPU and SYS \% CPU) for 30 million KV pairs. The graphs start with a straight line indicating the Write operation. The change in the slope of the line indicates that the write operation is completed and the Read operation has begun. This second part of the line indicates the CPU consumption during read operation. The series lines end at different times for different devices, indicating the faster completion time. With regards to \% CPU consumption, FS900 in asynchronous is found to be an efficient choice next only to RAM. Considering the fact that FS900 gives the peak performance with IO/s and OP/s, this minimal difference in CPU consumption can be eliminated. With \% SYS CPU consumption, the time spent on running kernel space system process, FS900 asynchronous with metadata cache enabled is found to be an optimal choice, especially for reads. The reason for the limited CPU consumption is due to the fact that, the total number of instructions per IO is reduced with CAPI. This leads to significant improvement in processor time spent in managing the IO. This, in effect, frees up processing resources for actual compute work and not just moving around data \cite{SolutionRefGuide}.

\subsection{Performance of \texttt{ark\_nextN} API }\label{sec:arknextN}

A normal read (without NextN) calls \verb|ark_get| API for each key in the ArkDB to return the value for the key, and these operations are done in parallel based on QD. Whereas, the \verb|ark_nextN| API  returns each key and not its value in the ArkDB, in a psuedo-random order, one operation at a time. The \verb|ark_next| API reads one KV pair at a time for each 4K reads. \verb|ark_nextN| API gets ``N'' keys at a time for each 4K reads in the ArkDB store. For the dataset in this study: key=8 Bytes and values =24 Bytes. This is about 100x faster for just reading the keys.

% Table generated by Excel2LaTeX'
\begin{table}[htbp]
  \centering
  \caption{Performance of ark\_ nextN API for SSD, NVM, RAM, FS900 in Asynchronous Mode for 100 million (5 GB) KV dataset}\label{tab:getN}
    \begin{tabular}{clccccrc}
    \hline
    \multirow{3}[0]{*}{Device} & \multicolumn{1}{c}{\multirow{3}[0]{*}{G Value}} & \multicolumn{3}{c}{Write} & \multicolumn{3}{c}{Read} \\
    \cmidrule(lr){3-8}
          &       & \multirow{2}[0]{*}{OP/s} & \multirow{2}[0]{*}{IO/s} & \multirow{2}[0]{*}{Time } 
                  & \multirow{2}[0]{*}{OP/s} & \multirow{2}[0]{*}{IO/s} & \multirow{2}[0]{*}{Time } \\  
          
          &       &       &       &   (seconds)    &       &       & (seconds) \\
          \hline
          &       &       &       &          &       &       &                  \\
    \multirow{3}[0]{*}{SSD} 
          & 1 & 39,840 & 326,236 & 251   & 6,086 & 48,861 & 1,643 \\
          & 1000 & 40,000 & 356,924 & 250   & 5,250 & 42,148 & 2 \\
          &       &       &       &          &       &       &                  \\
    \multirow{3}[0]{*}{NVM} 
          & 1 & 41,152 & 367,205 & 243   & 6,644 & 53,341 & 1,505 \\
          & 1000 & 41,152 & 367,205 & 243   & 5,250 & 42,148 & 2 \\
          &       &       &       &          &       &       &                  \\
    \multicolumn{1}{c}{\multirow{3}[0]{*}{FS900}} 
          & 1 & 285,714 & 727,040 & 35    & 4,411 & 7,832 & 2,267 \\
          & 1000 & 294,117 & 748,424 & 34    & 3,973 & 5,772 & 23 \\
          &       &       &       &          &       &       &                  \\
    \multirow{3}[0]{*}{RAM}
          & 1 & 588,235 & 1,496,848 & 17    & 86,956 & 154,396 & 115 \\
          & 1000 & 555,555 & 1,413,689 & 18    & 91,385 & 132,769 & 1 \\
          \hline
    \end{tabular}%
\end{table}% 

With G=1 vs G=1000, there was around 800x improvement on read time from SSD and NVM and 100x improvement for FS900 and RAM. Table \ref{tab:getN} also indicates that, even though the read IO/s and OP/s was not the maximum or the optimal for the device. we were able to reduce the time-taken by a huge margin with \verb|ark_nextN| API. This was also the reason why we focused on getting the peak IO/s and OP/s rather than the time taken to completion.

\subsection{Large Experiments Results}\label{sec:LargeEx}

Using a CAPI Interface, FS900 performs better for read operations than the other devices,  including RAM. This becomes highly advantageous when the dataset size is very large. Table \ref{tab:LargeExRes}  shows the results of the performance experiments on larger datasets (120GB, 954GB, 1500GB and 3TB). The OP/s and IO/s for FS900 were almost consistent, irrespective of the dataset size. In the case of RAM, the performance of read/write OP/s and IO/s increased by around 1.5 times until the dataset size is less than the size of RAM ( 1 TB). When the dataset size gets larger than the size of RAM, RAM can have cache misses and the virtual memory manager would be using the system disk for swapping, thereby increasing the paging and latency. At this stage, we observed a drastic decrease in the performance of IO/s and OP/s for RAM. For the dataset with 100 billion KV pairs (3.17 TB), FS900 provided a steady 380K IO/s for 71 hours. 

% Table generated by Excel2LaTeX '
\begin{table}[htbp]
  \centering
  \caption{Large Experiments Results}\label{tab:LargeExRes}
    \begin{tabular}{ccrrrrrr}
    \hline
    \multirow{3}[0]{*}{Device} & \multicolumn{1}{r}{\multirow{3}[0]{*}{Dataset Size}} & \multicolumn{3}{c}{Write} & \multicolumn{3}{c}{Read} \\
    \cmidrule(lr){3-8}
          &       & \multirow{2}[0]{*}{OP/s} & \multirow{2}[0]{*}{IO/s} & \multirow{2}[0]{*}{Time} & \multirow{2}[0]{*}{OP/s} & \multirow{2}[0]{*}{IO/s} & \multirow{2}[0]{*}{Time} \\
          &       &       &       &   (seconds)    &       &       & (seconds) \\
          \hline
          &       &       &       &       &       &       &  \\
    \multirow{4}[0]{*}{FS900} & 120 GB & 166,223 & 331,259 & 18,048 & 358,808 & 381,680 & 8,361 \\
          & 954 GB & 179,750 & 357,695 & 138,386 & 383,222 & 383,222 & 64,910 \\
          & 1.5 TB & 188,628 & 375,354 & 210,148 & 383,822 & 383,822 & 103,277 \\
          & 3.17 TB & 189,564 & 377,477 & 515,391 & 382,732 & 382,732 & 255,270 \\
	     &   &  &   &   &   &   &   \\
    \multirow{3}[0]{*}{RAM} & 120 GB & 2,941,176 & 5,861,339 & 1,020 & 4,716,981 & 5,017,660 & 636 \\
          & 954 GB & 3,539,413 & 7,043,255 & 7,028 & 5,327,693 & 5,327,693 & 4,669 \\
          & 1.5 TB & 2,149,267 & 4,283,774 & 16,936 & 2,576,991 & 2,576,991 & 14,125 \\
    \hline
    \end{tabular}%
  %\label{tab:addlabel}%
\end{table}% 

In general, RAM would be faster for most of the operations, but it comes with a limited capacity (1TB). FS900 has 20TB of storage and can also persist the data. The effectiveness of FS900 with CAPI comes into picture only when the dataset size is very large compared to the size of RAM. The CAPI 2.0  in Power9 makes use of the PCIe Gen4 and provides double the performance.

\section{Related Works}\label{sec:relatedworks}
There are various studies on benchmarking and evaluating heterogeneous storage systems. Apache Cassandra Optimized durable commit log using CAPI-Flash \cite{sendir2016optimized} focuses on durable logging on flash using Power8 CAPI-Flash. Their CAPIFlash commit log showed a 107\% better throughput in write-only workloads compared to Cassandra’s durable alternative. 

Our work is the first to study the performance of the FS900 with CAPI accelerator card through CAPIFlash read write benchmark. We also evaluated the performance by varying the threads and other parameters of the CAPIFlash library and compared it with RAM, SSD and NVM without using the CAPI accelerator. We evaluated the read time / write time, OP/s and IO/s of RAM, SSD, NVM and FS900 with both synchronous and asynchronous modes. We also found the optimal CPU threads and other parameters of the CAPIFlash library to attain a peak performance. These results for different memory devices, and their communication modes when using CAPIFlash library, will be helpful for developers in estimating their applications design choices.

\section{Conclusions}\label{sec:conclusions}

In this paper, we examined the performance of the FS900 with CAPI accelerator card through a basic read/write benchmark and compared it to various heterogeneous storage devices. The asynchronous mode on all devices gave the best results with IBM FS900, giving the highest read IO/s. We were able to achieve peak performance on FS900 with CAPI, using a lesser number of threads. This enables the CPU resources to be efficiently used in other areas of the application. When the dataset size exceeds the capacity of RAM, FS900 would be a cost-effective alternative. 

The flash memory in the IBM FS900 has higher performance due to hardware only data path, its distributed random-access memory and its allowance of massive parallelism in handling the data.  The CAPI accelerator allows access to shared memory regions and cache areas as though they were a processor in the system. The  CPU overhead in managing the IO request is now offloaded to the CAPI accelerator. The data transfer operation minimizes thousands of instructions and requires much fewer clock cycles in the processor. This improves the read/write performance of FS900 with CAPI unlike standard IO-attached flash storage. CAPI enables the class of IO-bound applications to perform better, where these advantages are a critical factor. Examples include weather prediction modeling and cybersecurity.

In future work, we plan to evaluate the FlashSystem 9150 utilizing the openCAPI 3.1 on a Power 9+ architecture. We also hope to extend these studies to develop a real-time streaming grid application on the CAPI accelerator card using NASA's MODIS dataset.

\section*{Acknowledgement}
We would like to thank Mike Vageline of IBM Cognitive Systems and Software Development for his support on the CAPIFlash - The IBM Data Engine for NoSQL library. We wish to acknowledge the NASA GSFC Distribution Active Archive Center (DAAC) for providing the MODIS Surface Reflectance (MOD09) data acquired from the Level-1 and Atmospheric Archive and Distribution System (LAADS) used for this study. We wish to thank Dale Pearson of IBM Yorktown Heights for providing this unique Power 8+ configuration with the FlashSystem 900. Finally, we wish to acknowledge the NSF Center for Accelerated Real-Time Analytics  (NFS Award Number 1747724) and its industrial members for providing the resources to carry out this study.

% ---- Bibliography ----
%

\end{document}